\documentclass{article}

\usepackage{arxiv}

\usepackage{url}            % simple URL typesetting
\usepackage{graphicx}
\usepackage{float}

\usepackage[english]{babel}
\usepackage[square,numbers]{natbib}
\title{Provably safe systems: \\
    the only path to controllable AGI}

\author{ {\includegraphics[scale=0.06]{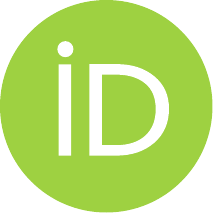}\hspace{1mm}Max Tegmark} \\
	Department of Physics\\
	Insitute for AI \& Fundamental Interactions\\
	Massachusetts Institute of Technology\\
	Cambridge, MA 02139 \\
	\And
        {\includegraphics[scale=0.06]{orcid.pdf}\hspace{1mm}Steve Omohundro} \\
	Beneficial AI Research\\
	Palo Alto, CA 94301\\
}

\renewcommand{\headeright}{}
\renewcommand{\undertitle}{}
\renewcommand{\shorttitle}{}

\begin{document}
\maketitle

\begin{abstract}
	We describe a path to humanity safely thriving with powerful Artificial General Intelligences (AGIs) by building them to provably satisfy human-specified requirements. We argue that this will soon be technically feasible using advanced AI for formal verification and mechanistic interpretability. We further argue that it is the only path which guarantees safe controlled AGI. We end with a list of challenge problems whose solution would contribute to this positive outcome and invite readers to join in this work.
\end{abstract}

% keywords can be removed

\section{Introduction}
\begin{center}
\textit{``Once the machine thinking method had started, \\
it would not take long to outstrip our feeble powers. \\
At some stage therefore we should have to expect the machines to take control''\\}
Alan Turing 1951 \citep{TuringQuote}
\end{center}

AGI \citep{WikipediaAGI} safety is of the utmost urgency, since corporations and research labs are racing to build AGI despite prominent AI researchers and business leaders warning that it may lead to human extinction \citep{BibEntry2023Sep}. While governments are drafting AI regulations, there’s little indication that they will be sufficient to resist competitive pressures and prevent the creation of AGI. Median estimates on the forecasting platform Metaculus of the date of AGI's creation have plummeted over the past few years from many decades away to 2027 \citep{MetaculusWeakAGI} or 2032 \citep{MetaculusStrongAGI} depending on definitions, 
with superintelligence expected to follow a few years later \citep{MetaculusASI}.

Is Alan Turing correct that we now \textit{``have to expect the machines to take control''}? If AI safety research remains at current paltry levels, this seems likely. Considering the stakes, the AI safety effort is absurdly small in terms of both funding and the number of people. One analysis \citep{McAleese2023Jul} estimates that less than \$150 million will be spent on AI Safety research this year, while, for example, \$63 billion will be spent on cosmetic surgery \citep{CosmeticSurgery} and \$1 trillion on cigarettes \citep{CigaretteMarket}. Another analyst estimates \citep{BenjaminTodd} that only about one in a thousand AI researchers works on safety. 

Much of the current AI safety work is focused on ``alignment'' which attempts to fine-tune deep neural networks so that their behavior becomes more aligned with human preferences. While this is valuable, we believe it is inadequate for human safety, especially given the profusion of open-source AI that can be used maliciously. In the face of the possibility of human extinction, we must adopt a ``security mindset'' 
\citep{Schneier} and rapidly work to create designs which will be safe also against adversarial AGIs. With a security mindset, we must design safety both into AGIs and also into the physical, digital, and social infrastructure that they interact with \citep{Leveson}. AGI computations are only dangerous for us when they lead to harmful actions in the world.

In current approaches, many researchers seem resigned to having to treat AGIs as inscrutable black boxes. Indeed, some younger researchers seem to subconsciously equate AI with neural networks, or even with transformers. This manifesto argues that such resignation is too pessimistic. Mechanistic interpretability is opening up the black box and, once opened, it can be distilled into formal representations that can be reasoned about. 

We argue that mathematical proof is humanity's most powerful tool for controlling AGIs. Regardless of how intelligent a system becomes, it cannot prove a mathematical falsehood or do what is provably impossible. And mathematical proofs are cheap to check with inexpensive, extremely reliable hardware. The behavior of physical, digital, and social systems can be precisely modeled as formal systems and precise ``guardrails'' can be defined that constrain what actions can occur. Even inscrutable AI systems can be required to provide safety proofs for their recommended actions. These proofs can be precisely validated independent of the alignment status of the AGIs which generated them. 

There is a critical distinction between an AGI world which turns out to be safe and an AGI world in which humanity has extremely high confidence of safety. If a mathematical proof of safety doesn't exist, then it is very likely that malicious AGIs will find the vulnerabilities and exploit them \citep{Apruzzese2022Dec}. If there is a proof of safety and it is made explicit and machine-checkable, then humanity can trust protected machines to design, implement, and execute safe actions. The design and creation of provably safe AGI and infrastructure is both extremely valuable and can become increasingly routine to implement, thanks to AI-powered advances in automated theorem proving and mechanistic interpretability.

\section{The path to provably safe AI}

Our approach to AI safety traces back to Max's 2018 \textit{IJCAI} plenary talk \textit{``Intelligible Intelligence''} \citep{IJCAI}, Steve's 2013 paper \textit{``Autonomous Technology and the greater human good''} \citep{Omohundro2014Jul} and his 2023 talk \textit{``Provably Safe AGI''} \citep{Omohundro2023May} at the \textit{2023 MIT Mechanistic Interpretability Conference}.

Nancy Leveson's excellent text \textit{``Engineering a Safer World''} \citep{Leveson} describes a \textit{systems} approach to safety design. It involves modeling the capabilities of the adversary, the nature of the potential harm, and the surrounding system which enables that harm. This is relevant both to currently ongoing AI harm and to upcoming existential threats.

The capabilities of today's AIs provide a lower bound on future adversarial AGI capabilities. Current \textit{Large Language Models} (LLMs) \citep{WikipediaLLM} can write code \citep{MetaCodeLlama}, prove theorems \citep{Lample2022May}, design molecules \citep{ProtGPT2}, reverse engineer \citep{Fraser2023Jan}, plan attacks \citep{Richard2023Apr}, deceive \citep{Park2023Aug} and manipulate people \citep{Mindshift2023Apr}, read and reason about every human text, recording, and video, etc. These capabilities will only improve in coming years, and we should expect AGIs to eventually surpass the very best humans in each of these areas. For example, \textit{every} AGI will likely be able to hack systems and break codes at the level of the very best human hackers today.  

Several groups are working to identify the greatest human existential risks from AGI \citep{WikipediaAIRisk}. For example, the \textit{Center for AI Safety} recently published \textit{``An Overview of Catastrophic AI Risks''} \citep{Hendrycks2023Jun} which discusses a wide range of risks including bioterrorism, automated warfare, rogue power seeking AI, etc. Provably safe systems could counteract each of the risks they describe.

These authors describe a concrete bioterrorism scenario in section 2.4: a terrorist group wants to use AGI to release a deadly virus over a highly populated area. They use an AGI to design the DNA and shell of a pathogenic virus and the steps to manufacture it. They hire a chemistry lab to synthesize the DNA and  integrate it into the protein shell. They use AGI controlled drones to disperse the virus and social media AGIs to spread their message after the attack. Today, groups are working on mechanisms to prevent the synthesis of dangerous DNA \citep{Esvelt}. But provably safe infrastructure could stop this kind of attack at every stage: biochemical design AI would not synthesize designs unless they were provably safe for humans, data center GPUs would not execute AI programs unless they were certified safe, chip manufacturing plants would not sell GPUs without provable safety checks, DNA synthesis machines would not operate without a proof of safety, drone control systems would not allow drones to fly without proofs of safety, and armies of persuasive bots would not be able to manipulate media without proof of humanness.    

Let's examine the hardware, software, and social systems needed to provide those kinds of safety guarantees. Humanity's most powerful mechanism to control adversarial AGI is mathematical proof. Even the most powerful AGI can't fake safety by proving an incorrect theorem, and theorems can be cheaply and reliably checked. Proofs of safety can be verified in advance or in real time as actions are occurring. Let's define the terminology we will use.

A \textit{Provably Compliant System} (PCS) is a system (hardware, software, social or any combination thereof) that provably meets certain formal specifications.

\textit{Proof-carrying code} (PCC) \citep{Necula1999Jan} is software that not only is provably compliant, but that also carries within it a formal mathematical proof of its compliance, \textit{i.e.}, that executing it will satisfy certain formal specifications. Because of the dramatic improvements in hardware and machine learning \citep{Lample2022May} since George Necula first proposed PCC in his seminal 1998 thesis \citep{NeculaThesis}, it is now feasible to expand the scope of PCC far beyond its original applications such as type safety, since ML can discover proofs too complex for humans to create \citep{AIProofs}. Among others, Stuart Russell \citep{StuartRussell} has made a strong case for how PCC can advance AI safety. 

\textit{Provably Compliant Hardware} (PCH) is physical hardware whose operation is governed by a Provable Contract. 

\textit{Provable Contracts} (PC) govern physical actions by using secure hardware to provably check compliance with a formal specification before actions are taken. They are a generalization of blockchain \textit{"Smart Contracts"} \citep{SmartContract} which use the cryptographic guarantees of blockchains to ensure that specified code is correctly executed to enable blockchain transactions. Provable contracts can control the operation of devices such as drones, robots, GPUs and manufacturing centers. They can ensure safety by checking cryptographic signatures, zero-knowledge proofs, proof-carrying code proofs, etc. for compliance with the specified rules. 

\textit{Provable Meta-Contracts} (PMC) impose formal constraints on the creation or modification of other provable contracts. For example, they might precisely define a voting procedure for updating a contract. Or they might encode requirements that provable contracts obey local laws. At the highest level, a PMC might encode basic human values that all PCs must satisfy. 

Provably compliant systems form a natural hierarchy of software and hardware. If a GPU is PCH, then it should be unable to run anything but PCC meeting the GPU’s specifications. As far as software is concerned, PCH guarantees are analogous to immutable laws of physics: the hardware can’t run non-compliant code anymore than our physical universe can operate a perpetual motion machine or superluminal travel devices. Moreover, a PCC can be often be conveniently factored into a hierarchy of packages, subroutines and functions that have their own compliance proofs. So even if our universe were to turn out to be a computer simulation, superluminal travel would be provably impossible in our simulated world.

If a \textit{provable contract} controls the hardware that PCC attempts to run on, it must comply with the specification. Compliance is guaranteed not by fear of sanctions from a court, but because it is provably physically impossible for the system to violate the contract.

\begin{figure}[H]
    \centering
    \includegraphics[width=15cm]{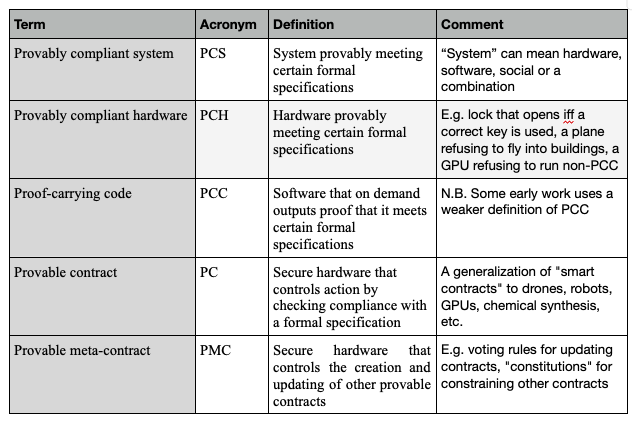}
\end{figure}

Controversially, we envision that powerful deployed PCCs will generally not be neural networks. We consider it likely that neural networks will help with knowledge and algorithm discovery, program synthesis and proof discovery,  but that PCCs will generally not be neural networks, because the inner workings of \textit{e.g.} language models appear far too messy to lend themselves to easy formal verification. Black box AI systems could either synthesize code directly (\textit{e.g.} like Code Llama \citep{MetaCodeLlama}), or be auto-converted to code though AI-powered mechanistic interpretability as described below.

Neural networks aren’t necessary for efficiently executing PCC, because there are plenty of efficient massively parallel computational architectures that can run proof-carrying code on today's and tomorrow's GPU server infrastructure. A possible path to safety is:

\begin{enumerate}
    \item Use AI to discover algorithms and knowledge
    \item Use AI to implement this as provably compliant proof-carrying code (PCC) \citep{WikipediaPCC}
    \item Create a global industry standard for provably compliant hardware (PCH) and operating systems (written in PCC) that only run PCC that self-proves compliance with risk-commensurate specifications
\end{enumerate}

The early focus should be on key components of systems with existential risks. In particular, we absolutely need to lock down nuclear weapons and biohazard laboratories from adversarial AGI attacks. Next on the list are the GPUs which run AGIs, the factories which manufacture them, and the supply chains which are essential to them. Control of this sector can be used to limit the rate of advancement of frontier AGI. 

We also see many benefits to using provable technology throughout society. It has the potential to solve many of today's social dilemmas and to create a more harmonious and productive world.   

\section{Mathematical Proof}

Mathematical proof is central to our approach to AGI safety. Its current state is both highly developed and rather chaotic. The development and clarification of mathematical proof is one of humanity's greatest accomplishments \citep{Stillwell2022Nov}. Starting around 350BCE, thinkers like Aristotle \citep{WikipediaAristotle} and Euclid \citep{WikipediaEuclid} sought to systematize mathematical thinking. By 1714, Leibniz dreamed of a ``sp\'ecieuse g\'en\'erale'', in which all truths of reason would be reduced to a kind of calculus" \citep{Leibniz}. By 1854, Boole, Cantor, and Frege had created the foundations of modern logic. And in 1925, the full foundations of modern mathematics were created in the form of ``Zermelo-Frankel Set Theory'' (ZFC) \citep{WikipediaZFC}. This system can encode all of mainstream mathematics, physics, engineering, computer science, economics, and other precise disciplines and has easy to check proofs.

Mathematical proof has been intertwined with computation throughout its development. In 1936, Alan Turing created a precise model of computation \citep{WikipediaTuring} and showed how to use mathematical proof to guarantee correctness of algorithms. As practical computers became available, they were used to prove theorems in propositional logic in 1956 and first order logic in 1976 \citep{WikipediaTheorem}. In the 1980s \citep{Boulanger2013May}, the field of ``Formal Methods'' \citep{WikipediaFormal} arose to prove properties of hardware and software. By 2000, propositional theorem provers (SAT/SMT solvers \citep{Committee2023Apr}) were routinely used to check chip designs and critical software components. But until recently, systems were not able to fully prove important theorems in ZFC. A series of ``Proof Assistants'' arose (eg. HOL, Mizar, MetaMath, Coq, Lean, Isabelle, etc. \citep{WikipediaProof}) in which weak theorem provers assisted human mathematicians in creating proofs. As of 2023, MetaMath has over 23,000 \citep{Metamath} and Lean has over 100,000 machine-checked human-proven theorems \citep{WikipediaLean}.

In 2020, OpenAI used transformer-based GPT-f to automatically prove 56.5\% of MetaMath theorems \citep{Polu2020Sep}. In 2022, a group at Meta used \textit{``HyperTree Proof Search''} to prove 82.6\% of MetaMath theorems \citep{Lample2022May}. Several groups are extending these results to Lean and other systems. Various ``Autoformalization'' efforts are also underway to automatically convert informal natural language specifications and proofs into formal ones \citep{Wu2022May}. These developments form the foundations of provably safe systems. Current AI does not yet have all of the capabilities we require, but trends suggest that it soon will. 

\section{Proof-Carrying Code}

Proof-carrying code is a fundamental component in our approach. Developing it involves four basic challenges:

\begin{enumerate}
    \item Discovering the required algorithms and knowledge
    \item Creating the specification that generated code must satisfy
    \item Generating code which meets the desired specification
    \item Generating a proof that the generated code meets the specification
\end{enumerate}

Generating a proof that the generated code meets the specification

Before worrying about how to formally specify complex requirements such as ``don’t drive humanity extinct'', it’s worth noting that there’s a large suite of unsolved yet easier and very well-specified challenges whose solution would be highly valuable to society and in many cases also help with AI safety. Here are a few examples:

\textbf{Provable cybersecurity:} One of the paths to AI disaster involves malicious use, so guaranteeing that malicious outsiders can’t hack into computers to steal or exploit powerful AI systems is valuable for AI safety. Yet embarrassing security flaws keep being discovered, even in fundamental components such as the ssh Secure Shell \citep{WikipediaHeartbleed} and the bash Linux shell \citep{WikipediaShellshock}. It’s quite easy to write a formal specification stating that it’s impossible to gain access to a computer without valid credentials.

\textbf{Securing the blockchain:} Attempts to formally verify \textit{e.g.} the Ethereum blockchain are making significant progress \citep{Ethereum}, but are still not complete. Preventing future AI from hacking and stealing cryptocurrency on a massive scale, not to mention wrecking other blockchain-based systems,  would contribute to AI safety. 

\textbf{Securing privacy:} Preventing AI from cracking privacy protocols and accessing secrets that humanity is trying to keep from AI or malicious actors would also be progress toward AI safety. This includes cryptosystems, end-to-end encrypted communication protocols and their surrounding infrastructure for password resets \citep{Binder2022Aug}, device migration \citep{One2017Mar}, etc. PCC combined with zero-knowledge proofs \citep{ZK} can enable a whole new spectrum of bespoke privacy possibilities \citep{Moocs2023Apr}, \textit{e.g.} provable guarantees about which information is revealed under which circumstances.

\textbf{Securing critical infrastructure:} Formal verification can also help secure critical infrastructure more generally, by ensuring that not only individual systems but also their integration is safe. Critical infrastructure needs to be PCH controlled by PCC. It is not difficult to imagine scenarios where an AGI hacking critical infrastructure could cause disaster, and humanity’s track record on critical infrastructure security leaves much room for improvement. For example, for 15 years, the code used to prevent an unauthorized launch of US nuclear missiles was ``00000000'' \citep{Nukes}.

These examples show that a major effort toward requiring AI to synthesize proof-carrying code can produce immediate improvements in AI safety (and provide great societal value more generally), while at the same time taking valuable steps toward solving the proof-carrying AGI challenge. 

Proof-carrying AGI running on PCH appears to be the only hope for a guaranteed solution to the control problem: no matter how superintelligent an AI is, it can’t do what’s provably impossible. So, if a person or organization wants to be sure that their AGI never lies, never escapes and never invents bioweapons, they need to impose those requirements and never run versions that don’t provably obey them. 

Proof-carrying AGI and PCH can also eliminate misuse. No malicious user can coax an AGI controlled via an API to do something harmful that it provably cannot do. And malicious users can’t use an open-sourced AGI to do something harmful that violates the PCH specifications of the hardware it must run on. There must be global industry standards that check proofs to constrain what code powerful hardware and operating systems will run.

In terms of policy, AI would become more like biotech is today. Before a biotech company is allowed to market a new drug, they need to prove to government-appointed experts at e.g. the FDA that it satisfies certain safety standards. Analogously, before a potentially harmful AI is allowed to be deployed, it would need to prove that it satisfies certain safety standards. In both the biotech and AI cases, these safety standards would be agreed upon by society, but in the AI case, the enforcement would be fully automated: all providers of sufficiently powerful hardware would be mandated to ensure that their hardware would not run code which doesn't carry a proof of compliance with these standards. Just as the existence of the FDA incentivizes massive industry investments in clinical trials and other safety research, such AI regulations would incentivize (sorely needed!) massive industry investments in AI safety, placing the responsibility where it belongs: on those selling and deploying the technology.

AI safety policy would also become more like biotech safety by including \textit{physical} security requirements, which would be formally specified as PCH. That’s how we prevent rogue DNA synthesis, rogue GPU creation, rogue robot attacks, rogue system self-replication, etc.

\section{Automated Software Verification}

Let's examine where we are on each of the four steps to PCC listed above:

\begin{enumerate}
    \item Discovering the required algorithms and knowledge
    \item Creating the specification that generated code must satisfy
    \item Generating code which meets the desired specification
    \item Generating a proof that the generated code meets the specification
\end{enumerate}

Traditionally, all four steps have been performed by humans, but for many tasks, ML is now better than humans at Step 1. For example, machine-learned translators, image generation algorithms or GPT4-style chatbots outperform anything humans know how to code. But with current LLM techniques, this highly valuable algorithmic knowledge is encoded in an inscrutable black-box form. 

In our approach to AI safety, we \textit{want} to keep Step 2 in the hands of humans, but assisted by AGI to ensure that all desirable properties are fully specified. Currently, the two really hard steps are 3 and 4. With gargantuan effort, humans have been able to create entire formal operating systems (\textit{e.g.} the seL4 Microkernel \citep{seL4Foundation2023Sep}) together with formal proofs of its properties. But for provably safe infrastructure to solve the AGI safety problem, we need AI to build these systems rapidly and at scale. It is possible to first generate code and then to generate proofs of its properties, but it is often better to create the code and the proof together.

Automated theorem proving \citep{WikipediaAutomated} and formal verification \citep{WikipediaFormal} have a long and successful history, in which most of the proofs were created by humans, often supported by automated assistants in a human-machine collaboration \citep{Amrani2018Jun}. A great example is Dafny \citep{Dafnyproject2023Jul}, a programming and verification language that aims to formally verify that the user-provided code satisfies user-provided specifications. Although the verification process is partly automated, users need to help it along in complex situations by providing hints like pre-and-post conditions, loop invariants and termination conditions. However, we may be on the cusp of a revolution here, with large language models and other AI tools showing signs of catching up with and overtaking humans in theorem-proving ability. If current trends continue, then just as AI has already overtaken human ability in so many other domains, we may soon routinely use AI to prove program properties vastly faster and better than is possible today.

Abstractly, a proof is simply a string of symbols, so the number of candidate proofs grows exponentially with the maximum length allowed. Although it can be notoriously difficult to find proofs, they are fast and simple to verify that you have found a valid one. MetaMath has a simple and efficient proof verifier that’s only 300 lines of Python \citep{Meta300}, runs in linear time, and can check arbitrary ZFC proofs. This means that formal verification is fundamentally a \textbf{search problem}: you’re searching an exponentially large space of proof strings, but once you find a valid proof, it’s trivial to realize that you’ve found one. 

Modern AI techniques have already revolutionized other exponential search problems, for example searching through the exponentially large sets of future game rollouts to win at Go and Chess with DeepMind’s AlphaZero \citep{AlphaZero}. For propositional logic theorem proving, SAT-solvers such as such as Microsoft’s Z3 \citep{Z3} do an excellent job on practical problems of searching through the exponentially large set of possible solutions. As mentioned above, deep learning theorem provers can already prove 82.6\% of MetaMath's theorems with no human input whatsoever. Large Language Models (LLMs) \citep{WikipediaLLM} may be able to dramatically improve this performance by using a more human-like top-down approach to guiding the search, splitting the proof goal into natural informal subgoals \citep{Zhao2023May} that can be tackled separately. Since the search space size grows exponentially with the proof length n, such a divide-and-conquer approach of proving shorter subgoal proofs separately can offer  exponential speedup: $2^n + 2^n$ is exponentially smaller than $2^{2n}$. Another reason to be bullish on ML for theorem proving is that autoformalization \citep{Szegedy2020Jul} and synthetically generated examples are likely to soon produce massive training datasets.

We can look at human programmers to see why decomposition of a proof task into smaller informal subgoals will often be successful. Good human programmers always have an argument for why the code they are writing is correct. They generally split large programs into smaller components and create mental arguments for the correctness of each component and of their combination into the large program. Formalizing this intuitive multi-step proof will greatly accelerate the search for a rigorous proof. Large human-written code tends to be organized into many smaller functions and subroutines (reflecting the modular nature of the algorithmic tasks the program accomplishes) that can be verified separately. AIs will be able to generate and access large libraries of subroutines together with proofs of their formal properties. In summary, we still lack fully automated code-verifying AI powerful enough for our provably safe AI vision, but given the rate AI is advancing, we are hopeful that it will soon be available. Indeed, just as several other AI fields dominated by GOFAI (``Good Old-Fashioned AI'') techniques were ripe for transformation by machine learning around 2014, we suspect that automated theorem proving is in that pre-revolution stage right now.

It’s important to emphasize that formal verification must be done with a security mindset, since it must provide safety against all actions by even a superintelligent adversary. Fortunately, the theoretical cryptography community has built a great conceptual apparatus for digital cryptography. For example, Boneh and Shoup's excellent new text \textit{``A Graduate Course in Applied Cryptography''} \citep{Boneh} provides many examples of formalizing adversarial situations and proving security properties of cryptographic algorithms. But this security mindset urgently needs to be extended to hardware security as well, to form the foundation of PCH. As the lock-picking lawyer \citep{LockPickingLawyer2021Apr} quips: ``Security is only as good as its weakest link'' \citep{Cuthbertson2019Aug}. For physical security to withstand a superintelligent adversary, it needs to be provably secure. 

\section{Program Synthesis and Mechinterp}

\begin{figure}
	\centering
        \includegraphics[width=15cm]{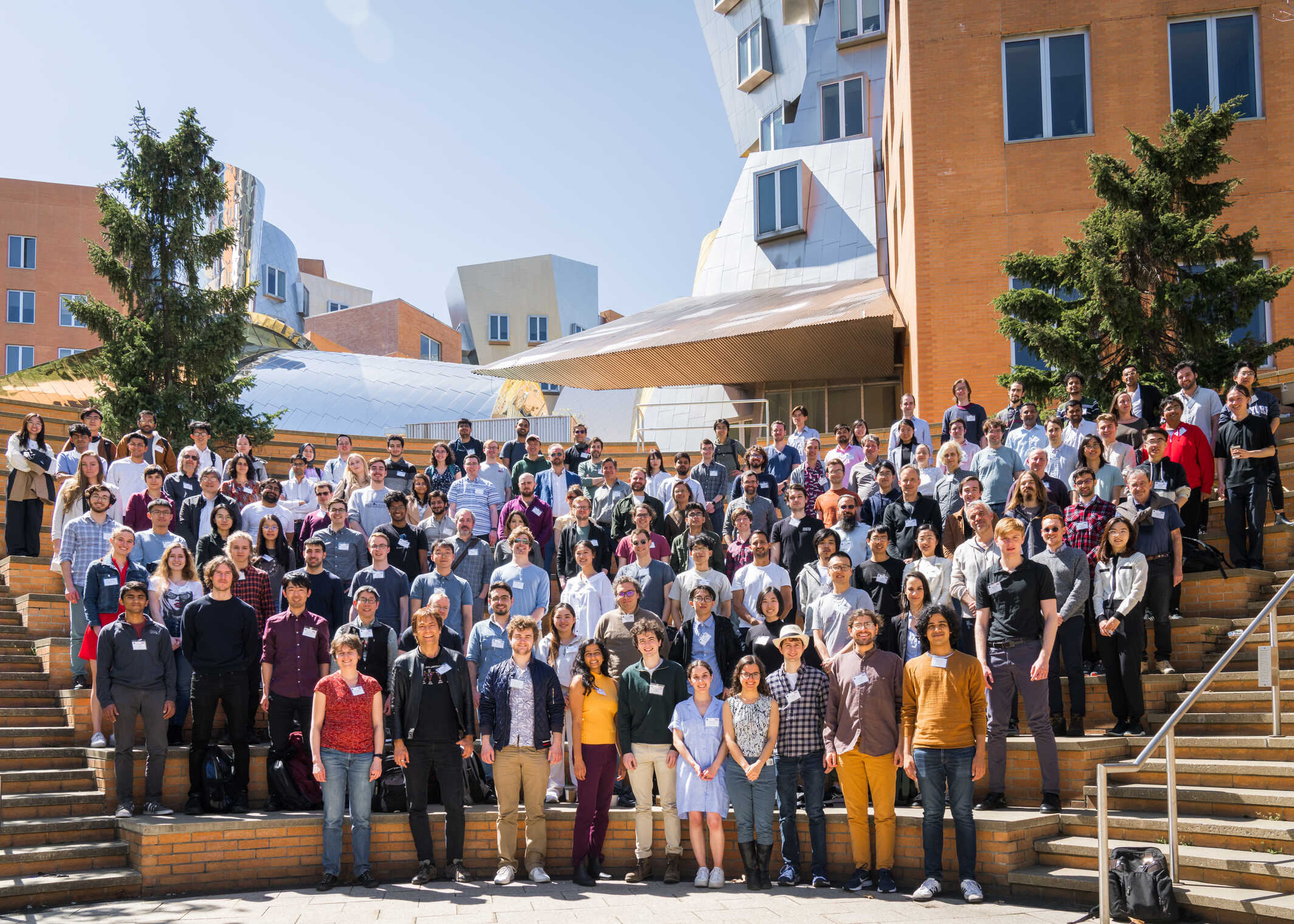}
	\caption{The 2023 MIT Mechanistic Interpretablity Conference}
\end{figure}

Finally, we consider Step 3: implementing an AI's inference not as a black box neural network, but as code which can be formally verified. First, note that the seemingly magical power of neural networks  comes \textbf{not from their ability to execute, but from their ability to learn}. Once training is complete and it is time for execution, a neural network is merely one particular massively parallel computational architecture – and there are many others based on traditional software that can be similarly efficient. 

But neural networks routinely crush the competition when it comes to \textit{learning}. They’re not only computationally \textit{universal}, but they’re also \textit{differentiable} with respect to their parameters. This means that learning reduces to a continuous minimization problem with gradients that provide information on how to improve on each step. In contrast, searching the space of all possible Python programs to minimize a loss function involves a daunting discrete combinatorial search over an exponentially large space. One might summarize that the remarkable power of neural networks comes from their \textbf{differentiability}, not from their \textbf{inscrutability}. The black box helps for learning, not for execution. If the provably safe AI vision succeeds by replacing powerful neural networks by verified traditional software that replicates their functionality, we shouldn’t expect to suffer a performance hit. Once a good algorithm has been discovered, neural networks lose their computational advantage. Indeed, PCCs may be more efficient, because neural networks have been shown to underperform GOFAI methods on basic tasks such as low-dimensional interpolation \citep{Michaud2022Oct}.

There are two strategies for using AI to produce the PCCs needed in Step 3. First, we can use LLMs like Meta's ``Code Llama'' \citep{MetaCodeLlama} to directly write the code (and eventually the proofs as well). These systems still sometimes generate incorrect code, but are good enough to be useful co-pilots for human programmers. Human brains are also inscrutable black-boxes, and yet human scientists can express how they perform certain tasks (\textit{e.g.} predict how objects move under gravity) in symbolic form which can be communicated to others and implemented as a computer program. 

Since we humans are the only species that can do this fairly well, it may unfortunately be the case that the level of intelligence needed to be able to convert all of one's own black-box knowledge into code has to be at least at AGI-level. This raises the concern that we can only count on this ``introspective'' AGI-safety strategy working after we’ve built AGI, when according to some researchers, it will already be too late. 

The second strategy is to delegate this task to another AI which has full access to the first AI’s trained neural network. We call this artificial neuroscience, since traditional neuroscientists study biological black-box brains and try to figure out how they do what they do. This AI subfield has seen rapid recent progress under the label \textit{``Mechanistic Interpretability''} \citep{Nanda2023Jan} (often shortened to \textit{``Mechinterp''} or \textit{``MI''}). It is still a very small field, as can be seen from Figure 1 which shows how few people there were at the its largest conference to date \citep{MIConf}.

Mechinterp is already achieving impressive results and aspires to three goals:

\begin{enumerate}
    \item Diagnose trustworthiness
    \item Improve trustworthiness    
    \item Guarantee trustworthiness
\end{enumerate}

The third goal will be accomplished if MI-tools can convert the knowledge and algorithms auto-discovered by a black-box NN into verified code.

The rate of MI progress is quite remarkable. In the last year, the field has made more fundamental progress than all of traditional neuroscience combined, despite having orders of magnitude fewer researchers. Talks at the conference \citep{MIConf} reported breakthroughs in understanding how LLMs represent knowledge and algorithms. LLMs aren’t irreducible blobs, but share recurrent high-level circuits and motifs and can sometimes be decomposed into modules (computational ``quanta''). In contrast to traditional neuroscience, artificial neuroscience has the advantage of access to every neuron's activity and synaptic weight. And MI-researchers can create and study simplified networks to understand general principles. Once powerful AI's are tasked with Mechinterp, we may see yet another ML-powered revolution.

\section{Networks of Provable Contracts}

Humanity's most urgent task is preventing existential risk. For example, we must protect high-risk facilities such as biohazard labs, nuclear weapons facilities, advanced drone manufacturing, etc. from malicious use by AGIs or AGI-enabled malicious humans. Today, the rules for launching a nuclear weapon are described in complex classified manuals and are implemented by military personnel. Unfortunately, it is likely that humans will be manipulable by AGIs via \textit{e.g.} persuasion, bribery or blackmail. 
Giving small groups of humans too much power makes them targets for attack. So it is risky if small groups
of humans can cause cause catastrophic damage. But we certainly don't want AGI to have such power either! 

Such failure modes can be avoided with provable contracts implemented in secure hardware, with precisely specified conditions enabling each possible action. Many groups are currently developing tamperproof chips \citep{WikipediaTamper} and shielded devices with sensors to detect attempted tampering. Cryptographic keys can be deleted when tampering is attempted (``Zeroisation'' \citep{WikipediaZeroisation}). Apple is including ``Secure Enclaves'' in all of their products \citep{Apple} (including the Apple watch) which are dedicated portions of their system chips that implement truly random number generation, secure key storage, fast hardware cryptographic encoding and decoding, and ability to securely store secret data in external memory. Intel provides similar facilities in their ``Software Guard Extensions'' (SGX) \citep{IntelSGX}. 

Guaranteed security against adversarial AGI requires the aforementioned Provably Compliant Hardware (PCH), with precise proofs and mathematical models in the style of those used today in theoretical cryptography. We need precise formal models of hardware designs, the capabilities of an adversary, and desirable and unacceptable outcomes. Every sensitive device (eg. DNA synthesis machine, military drone, GPU creator, etc.) must be protected by a network of provable contracts implemented in PCH. Networks of devices can be better able to detect and prevent intrusion and to be robust in the face of attacks. 

The terms of contracts are essential both to the security of the protected systems and to creating a high quality of life for humanity. Enormous work in the social design of contracts is necessary. There need to be mechanisms to create contracts that meet human needs and to update them via provable metacontracts when conditions change, to create positive human value.

\section{The Only Path To Controllable AGI}

We hope that this proposal is inspiring. But it is quite complex with many components, some of which don't yet exist. Why should humanity devote resources to making this happen? In this section, we argue that this approach is likely the \textit{only} way to avoid existential risk. 

The first step in this argument is based on G\"{o}del's Completeness Theorem \citep{WikipediaGodel}(not to be confused with his Incompleteness Theorem) which says that any statement which is true in all models has a proof. Since even our largest computers have only a finite number of states, this implies that if some safety property doesn't have a proof, then there \textit{must} be a way to violate it. Sufficiently powerful AGIs may well find that way. And if that AGI is malicious or controlled by a malicious human, then it \textit{will} exploit that flaw. So we absolutely need a technological base for which there exists a safety proof.

In theory, we might have a provably safe infrastructure without explicitly expressing or checking proofs. But in such an infrastructure, humans and other systems will have to make decisions \textit{without guarantees of safety}. They might have a \textit{belief} that a system is safe or have some \textit{evidence} that it is safe. But in an adversarial setting it is very easy to manipulate beliefs and to manufacture or manipulate evidence. The only absolutely trustable information comes from mathematical proof. Because of this, we believe it is worth a fair amount of inconvenience and possibly large amounts of expense for humanity to create infrastructure based on provable safety. The 2023 global nominal GDP is estimated to be \$105 trillion \citep{Rao2023Aug}. How much is it worth to ensure human survival? \$1 trillion? \$50 trillion? Beyond the abstract argument for provable safety, we can consider explicit threats to see the need for it. 

Today, the top AI corporations are attempting to constrain the behavior of their models through techniques like ``Reinforcement Learning from Human Feedback'' (RLHF) \citep{WikipediaRLHF}. This builds a model of human preferences for different data and uses it to tune a generative neural network. While this is valuable for increasing alignment with human values, it is inadequate in an adversarial context. For example, the paper ``Fundamental Limitations of Alignment in Large Language Models'' \citep{Wolf2023Apr}
shows that ``any alignment process that attenuates undesired behavior but does not remove it altogether, is not safe against adversarial prompting attacks.'' 

Even if the leading AI laboratories are extremely careful with how they create and release their models, there is a thriving open source AI community. When Meta allowed academic researchers access to a powerful language model, the weights were leaked \textit{within a week} \citep{Wolf2023Apr} and were then distributed on bit torrent. Another powerful open source ChatGPT-style model was quickly altered by an anonymous hacker into ``Chaos-GPT'' with a goal to ``Destroy Humanity'' \citep{Lanz2023May}. With today's infrastructure, parties with malicious intent can freely run open source AI models on whatever hardware they can get access to. Normal political or corporate processes cannot stop these problems. We need an infrastructure that provably protects humanity against harm.

Many worry that even if the immediate existential AGI threats are dealt with, AGI will still likely amplify the competition between different human groups for wealth and political and military power. One influential exposition of this force is Scott Alexander's ``Meditations on Moloch'' which personalizes the competitive force as an unstoppable creature \citep{Moloch}. Some argue that AGI will embody and amplify that competitive force. Many of the harms of this kind of competition arise from ``social dilemmas'' \citep{SocialDilemmas}, ``collective action problems'' \citep{WikipediaCollective}, or ``prisoner's dilemmas'' \citep{WikipediaPrisoners} in which individuals or groups are incentivized to take selfish actions when they would be better off cooperating. Humans have moral emotions which encourage individuals to consider the impact of their actions on others. We have also created social mechanisms such as laws, police forces, and the legal system to disincentivize harmful actions. AGI need not be constrained by either of these forces. But we can construct networks of provable contracts which expose each participant to the externalities caused by their actions. Those have the potential to completely eliminate social dilemmas and to encourage win-win thinking and actions. 

In summary, we’ve described a vision for how humans can control AGI and superintelligence, where the only AGI that ever gets deployed consists of proof-carrying code. AI is allowed to write the code and proof, but not the proof-checker. The knowledge and algorithms used by the code can be AI-discovered, extracted either through self-explaining AI or with mechanistic interpretability tools. We’ve argued that this vision isn’t merely plausible, but that it may be the \textit{only} guaranteed solution to the control problem: no matter how superintelligent an AI is, it can’t do what is provably impossible. Losing control over superintelligent AI would arguably be humanity’s worst mistake ever, so proponents of alternative approaches to AGI safety owe a detailed explanation of how they guarantee success. For example, passing evaluations that screen for known risks are a necessary but not sufficient condition for safety. Absence of evidence of risk isn’t evidence of its absence. Mathematical proofs are!

\section{Challenge Problems}

In this appendix, we give examples of research directions that we encourage safety-concerned readers to tackle.  

\subsection{Automate formal verification:}

As described above, formal verification and automatic theorem proving more generally needs to be fully automated. The awe-inspiring potential of LLMs and other modern AI tools to help with this should be fully realized. 

\subsection{Develop verification benchmarks:}

The field of automatic theorem proving greatly benefits from the existence of large databases of machine-readable theorems and proofs for systems such as Metamath, HOL, Isabel and LEAN, and autoformalization \citep{Szegedy2020Jul} will hopefully expand them dramatically. It will be crucial for the formal verification field to develop an analogous large database of correct programs with formal specifications and corresponding compliance proofs.

\subsection{Develop probabilistic program verification:}

The formal version of neural learning and inference is ``probabilistic programming'' \citep{WikipediaPP}. It has recently been argued that probabilistic programming is a natural ``Language of Thought''\citep{Wong2023Jun} and that LLMs can naturally translate natural language into it, reasoning can be done in a rigorous Bayesian way, and the results translated back into language. The whole of Bayesian inference and learning can be naturally represented by probabilistic programs. We need formal verification for probabilistic programs that allow proofs of bounds on probability, that learning and inference were correctly performed on specified data, and that undesirable outcomes have probabilities below specified limits.

\subsection{Develop quantum formal verification:}

Develop a formalism that generalizes formal verification to quantum computing, the actual computational framework that our Universe appears to use.

\subsection{Automate mechanistic interpretability:}

As described above, we need to automate the task of extracting the knowledge and algorithms that black-box AI systems have learned during painstaking training. The field of mechanistic interpretability is making rapid progress, but this progress is still driven mainly by humans using hand-crafted techniques, and should be automated and powered by state-of-the-art AI tools.

\subsection{Develop mechanistic interpretability benchmarks:}

The field of mechanistic interpretability would greatly benefit from a large database of challenge problems with known answers. Such a database should contain an algorithm (say verified Dafny code implementing an algorithm to produce a training data set) as well as a neural network that has been trained on this dataset to perform the task well.  The task of MI-researchers is then to start only with the neural network (and optionally the training data), and automatically recover the algorithm or an equivalent one. As a simple example, the algorithm could be Quicksort, the dataset could be a spreadsheet with unsorted lists (input) and the sorted version (output), and the task would be to automatically extract whatever sorting algorithm the neural network has discovered.

\subsection{Build a framework for provably compliant hardware:}

Develop a rigorous yet practically useful formalism for provably compliant hardware, to match the existing framework for provably compliant software. Any physical system can be approximated as a classical or quantum computation, so a key challenge is to quantify and bound errors in such an approximation and translate them into practical bounds on behavior. Much of computer science rests on the assumption of substrate-independence (that the way information is processed can be mathematically described in a way that is independently of the hardware substrate  implementing the computation), and this idea needs to be generalized to all other relevant hardware, from locks and motors to DNA synthesizers. Ultimately, we need to unify formal methods for hardware,  software and mathematics. Formalizing the hierarchy of approximations used by physicists (\textit{e.g.} as presented in this excellent text: \citep{Thorne2017May}) would be a valuable step.

\subsection{Build a framework for provably compliant governance:}

AI safety involves not only hardware and software, but also society. To ensure human flourishing, we must not only ensure that machines treat humans well, but also ensure that humans, corporations and other organizations are given incentives that align with the well-being of others. This is a game theory problem of mechanism design, which we may term \textit{``Defeating Molloch''} \citep{Moloch}. It is so important that both Darwinian evolution and cultural evolution have developed many such collaboration mechanisms. For example, we evolved empathy and love, because groups with these traits out-competed those without. We invented gossip to disincentivize lying, cheating and freeloading. We invented a legal system to ensure that people and organizations acted for the greater good. Unfortunately, as adversaries get ever smarter, adversarial attacks against these mechanisms get ever more sophisticated. This process makes the adversarial security mindset presented here critical for humanity's future. In addition to the technological infrastructure, we also need to AGI-proof our social, legal and governance mechanisms. For example, the relatively dumb artificial intelligent agents we know as corporations have gotten increasingly successful at regulatory capture \citep{WikipediaRegulatory}, and no amount of technical AI safety work will suffice if the first tech company to build superintelligence can capture its national government and take over the world. Some argue that today's policymakers are already too cozy with the tech executives they are supposed to regulate. 

\subsection{Create provable formal models of tamper detection:}

The core of security in this system rests on the hardware that checks proofs. Ideally its design will be provably safe against specified levels of attack. Tamper detection \citep{WikipediaTamper} is critical for this. If an adversary expends enough energy, they are likely to be able to destroy any piece of physical hardware. But the overall security of the system must be resilient to that kind of attack. ``Zeroisation'' \citep{WikipediaZeroisation} erases cryptographic keys when tampering is detected. In some situations, there should also be a physical analog of zeroisation in which dangerous chemicals or biological materials are destroyed if intrusion is detected. Sensors and actuators might include fuses or other destructive elements which can be activated to incapacitate hardware. Such functionality  prevents an adversary from co-opting hardware, but risks becoming a denial-of-service vector. These destructive elements generally shift the game theoretic rewards away from physical attacks. For provable security, we'd like proofs of detection of \textit{every} physically possible attack. For example, a device might have sensors for temperature, electric fields, magnetic fields, electromagnetic radiation, acceleration, pressure, \textit{etc.} and a precise contract describing when and how to warn the broader network of an attack and when to delete potentially sensitive information. This approach should be fleshed out and made formal.

\subsection{Create provably valid sensors:}

Sensors are such a crucial type of hardware that they deserve special attention. Much of hardware security relies on sensors to detect adversaries, unexpected situations, or hardware modification, so we need high confidence in the accuracy of sensors. But a single video camera might be hacked and it's video stream altered. Multiple sensors can provide redundant information which can be checked for consistency. We need a formal framework for quantifying and increasing the reliability of sensor information. Ultimately, we'd like to have extremely high confidence in the state of a physical system which is attested to by cryptographic signatures and checkable proofs.

\subsection{Design for transparency:}

Much research has gone into developing good coding practices that facilitate verification.  Analogously, research is needed on how to design more easily verified hardware. One pernicious class of attacks on hardware involves introducing flaws that invalidate security properties. For example, this ``demonically clever'' analog attack on a chip introduces a hidden capacitor which is charged by a rare instruction and eventually opens a backdoor \citep{Greenberg2016Jun}. This kind of attack can be prevented by using provably protected manufacturing and provenance after manufacture. But it is desirable to be able to inspect this kind of hardware and to have provably high confidence of detecting dangerous flaws. We might call this \textit{``Design for Transparency''} in which available sensors are guaranteed to detect a certain level of attack to the physical structure.

\subsection{Create network robustness in the face of attacks:}

Von Neumann famously lectured on \textit{``Probabilistic Logics and the Synthesis of Reliable Organisms from Unreliable Components''} \citep{VonNeumann}. He adopted a probabilistic model of failure, but the general idea can be extended to adversarial attacks. We'd like a formal model of a network of provable contracts with precisely specified and proven resilience to physical and software attacks.

\subsection{Develop  useful applications of provably compliant systems:}

This will help motivate hard work on building PCS tools. Here are some examples worth fleshing out.

\subsubsection{Mortal AI: }

PCC could have a built-in expiration date after which it couldn’t run on PCH, which would include a PCH clock.

\subsubsection{Geofenced AI:}
 
PCH could have built rules for where it can operate, combined with a PCH GPS module measuring its location. This could provably guarantee non-proliferation and prevent various types of misuse.

\subsubsection{Throttled AI: }

By designing GPU’s as PCH that requires a ``crypto drip'' (a steady supply of crypto-tokens) to keep operating, it becomes easy to limit or shut down AI systems. The authors first learned of this idea from Anthony Aguirre \citep{aaguirre2022Jun}. This could be implemented hierarchically: for example, a regulatory agency could issue tokens to a safety-approved project, which would distribute them across its hardware infrastructure.

\subsubsection{AI kill switch: }

Such crypto-tokens could be made to expire after a short time, giving its issuer the power to shut down all AI systems that depended on them by halting the supply. This is analogous to how certain dangerous bacteria are bred to require a rare chemical that is absent in the environment, so that they promptly die if they escape the lab.

\subsubsection{Asimov-style laws: }

Many AI guiding principles have been proposed, from Asimov’s 3 Laws of Robotics to the Asilomar AI Principles \citep{Asilomar} and OECD AI Principles \citep{OECD}. What is the complete set of principles that can and should be made precise enough to be be used as PCS specifications?

\subsubsection{Least privilege guarantee: }

The ``Principle of Least Priviledge'' (PoLP) \citep{WikipediaLeast} requires that every agent (such as a process, user or a program) should have access to only the information and resources that are necessary for its legitimate purpose. Although the PoLP is a cornerstone of computer security, it is routinely flaunted. Can the PoLP be expressed as a formal specification that can be required by certain PCH? In contrast, some AI systems today are being given virtually unrestricted access to the world.

\section{FAQ: }

Q: Won’t debugging and ``evals'' \citep{Evals} guarantee AGI safety?
A: No, debugging and other evaluations looking for problems provide \textit{necessary} but not \textit{sufficient} conditions for safety. In other words, they can prove the presence of problems, but not the absence of problems.

Q: Isn’t it unrealistic that humans would understand verification proofs of systems as complicated as large language models?
A: Yes, but that’s not necessary. We only need to understand the specs and the proof verifier, so that we trust that the proof-carrying AI will obey the specs.

Q: Isn’t it unrealistic that we’d be able to prove things about very powerful and complex AI systems?
A: Yes, but we don’t need to. We let a powerful AI discover the proof for us. It’s much harder to discover a proof than to verify it. The verifier can be just a few hundred lines of human-written code \citep{Meta300}. So humans don’t need to discover the proof, understand it or verify it. They merely need to understand the simple verifier code. 

Q: Won’t checking the proof in PCC cause a performance hit?
A: No, because a PCC-compliant operating system could implement a cache system where it remembers which proofs it has checked, thus needing to verify each code only the very first time it’s used.

Q: Won’t it take decades to fully automate program synthesis and program verification? 
A: Just a few years ago, most AI researchers thought it would take decades to accomplish what GPT4 does, so it’s unreasonable to dismiss imminent ML-powered synthesis and verification breakthroughs as impossible. 

Q: Isn't it premature to work on provable safety before we know how to formally specify concepts such as ``Don't harm humans''?
A: No, because provable safety can score huge wins for AI safety even from things that are easy to specify, involving e.g. cybersecurity. 

\section{Acknowledgements}

Steve would like to thank Anthony Aguirre, James Barrat, Brad Cottel, Andrew Critch, David Dalrymple, Danit Gal, David Krueger, Ramana Kumar, Eric Rogstad, Jaan Tallinn, Jessica Taylor, Michael Vassar, Eliezer Yudkowsky, and Alex Zhu for helpful discussions of these issues.

\bibliography{provable}  %%% Uncomment this line and comment out the ``thebibliography'' section below to use the external .bib file (using bibtex) .

\end{document}